\RequirePackage{fix-cm}
\documentclass[twocolumn]{svjour3}          

\usepackage[english]{babel}
\usepackage{amssymb}
\usepackage{amsmath}

\usepackage{graphicx}
\usepackage{verbatim}
\usepackage{hyperref}
\usepackage{amsmath}
\newtheorem{defini}{Def.}

\begin{document}

\title{Gravity as field - field oriented framework reproducing General Relativity}
\subtitle{}

\author{Piotr Ogonowski         \and
        Piotr Skindzier
}

\institute{\and Leon Kozminski Academy,  ul. Jagiellonska 57, 03-301 Warsaw, Poland 
              \email{piotrogonowski.pl@gmail.com}     \\      
           \and
           WFAiS Jagiellonian University, ul. Reymonta 4, 30-059 Krakow, Poland 
              \email{piotr.skindzier@uj.edu.pl}
}

\maketitle

\begin{abstract}
In the last article we have created foundations for gravitational field oriented framework (DaF) that reproduces GR. In this article we show, that using
DaF approach, we can reproduce Schwarzschild solution with orbit equations, effective potential and constants of motion. Next we generalize results to other GR solutions and show, how gravitational field affects spacetime curvature and intrinsic spin of the bodies. It also appears, that field oriented approach requests to assign some spin value to the massless particles. Derived DaF framework has therefore significant meaning for searching for field based interpretation of gravity requested by quantum gravity.
 
\end{abstract}

\section{Introduction}
In this paper we continue researching project and develop field-oriented DaF framework reproducing General Relativity.\\
~\\
We use Einstein summation convention: commas denote partial derivatives and semicolons denote covariant derivative. We choose metric signature $(+,-,-,-)$ and consider c=1 for all calculations. We use spherical coordinates $(t,r,\theta,\varphi)$ for all the article.

\subsection{Basic DaF definitions}

In the previous article \cite{nasz} we have shown, that for Schwarzschild case, curved spacetime is equivalent to flat manifold minimally coupled to some scalar field.
It was also confirmed in \cite{adm}\\
This scalar field changes the metric of the Stationary Killing Observer (SKO) - the observer located in the constant position with distance r to the source of gravity. \\
~\\
We take the scalar field in form of 
\begin{equation}
\frac{1}{\gamma_g} \equiv \sqrt{1-\frac{r_s}{r}} \equiv \frac{d\tau}{dt}
\end{equation}
where
\begin{itemize}
\item $r_s$ is for Schwarzschild radius
\item $d\tau$ is the SKO's proper time
\item $dt$ is coordinate time of the stationary observer in infinity
\end{itemize}
~\\
To consider move of the test bodies in SKO reference frame, we introduce
\begin{defini}
\mbox{}
\begin{itemize}
\item $ds$ is the proper time of some test particle
\item $\gamma_v \equiv d\tau/ds$ is the dilation factor of the test particle in the SKO's local reference frame 
\item $E_0 \equiv m_0c^2$ is rest mass/rest energy of the test particle
\item $E_0 \cdot \varepsilon$ is total energy of the test particle in reference frame of the stationary observer in infinity
\end{itemize}
\end{defini}
~\\
At first, let us show, that using DaF approach we may reproduce Schwarzschild results.\\
~\\
In the next section, we will explain why it works and make generalization.

\section{Reproducing Schwarzschild results with DaF}

In the SKO's reference frame, we define Lagrangian with no action in the form of
\begin{equation} L= E_0 \cdot \varepsilon \left (\frac {1}{\varepsilon \gamma_g} - \frac {1}{\gamma_v} \right ) = 0 \end{equation}
giving
\begin{equation} 
\gamma_v = \varepsilon \cdot \gamma_g \label{epsil}
\end{equation}
To simplify calculations we normalize Lagrangian (divide by $E_0 \, \varepsilon$) to the form of
\begin{equation} 
\ell = \frac {1}{\varepsilon \gamma_g} - \frac {1}{\gamma_v}   \label{lag_red}  
\end{equation}
~\\
Let us consider spherically symmetric case and a spacetime with metric tensor $g_{\mu\nu}$
\begin{equation}
g_{\mu\nu} \equiv  \begin{bmatrix}
\frac{1}{\gamma_g^2} & 0 & 0 & 0\\ 
0 & -1 & 0 & 0\\ 
0 & 0 & -r^2\frac{1}{\gamma_g^2} & 0\\ 
0 & 0 & 0 & -r^2\sin^2(\theta)\frac{1}{\gamma_g^2}
\end{bmatrix} \label{metTen}
\end{equation}
resulting with metric
\begin{equation} 
ds^2=dt^2\frac{1}{\gamma_g^2}-dr^2-r^2d\theta^2\frac{1}{\gamma_g^2}-r^2\sin^2(\theta)d\varphi^2\frac{1}{\gamma_g^2}
\end{equation}
Thanks to symmetry we may always take $\theta=\pi/2$ and eliminate $d\varphi$ component.\\
~\\
To simplify calculations we note SKO's proper time as $d\tau$ and we also define SKO's angle coordinate $d\phi$ as
\begin{equation} 
d\phi \equiv \frac{d\theta}{\gamma_g}  \label{SKOangle}
\end{equation} 
what let us note local SKO metric as
\begin{equation} 
ds^2=d\tau^2-dr^2-r^2d\phi^2
\end{equation}

\subsection{Orbit equation}

We note normalized Lagrangian (\ref{lag_red}) for test particle considered in SKO's reference frame as
\begin{equation} 
\ell =  \frac{1}{\varepsilon}\sqrt{1-\frac{r_s}{r}} -  \sqrt{1- r^2 \left ( \frac{d\phi}{d\tau} \right )^2 - \left ( \frac{dr}{d\tau} \right )^2} \label{lagSch}
\end{equation}
Calculating derivative of the Lagrangian in respect to the angle $\phi $ we see, that angular momentum is conserved, and for $d\phi/d\tau$  derivative normalized angular momentum value $\alpha$ is equal to
\begin{equation} 
\alpha \equiv r^2 \frac{d\phi}{d\tau} \gamma_v \label{angmom}
\end{equation}
From above, we may describe angular velocity
\begin{equation} 
\frac{d\phi}{ds} = \frac{\alpha}{r^2}  \label{angvel}
\end{equation} 
In the plane of rotation we may note equation for momentum as
\begin{equation} 
\gamma_v^2 = \beta_v^2 \gamma_v^2 +1 =  \frac{dr^2}{ds^2} + \frac{\alpha^2}{r^2} +1 \label{simporb}
\end{equation}
therefore
\begin{eqnarray}
&& \frac{dr}{ds} = \sqrt{\gamma_v^2 - \left ( 1+ \frac{\alpha^2}{r^2} \right )}  \\  
&& d\phi = \frac{\alpha \; dr}{r^2 \sqrt{\gamma_v^2 - \left ( 1+ \frac{\alpha^2}{r^2} \right )}} \label{eq:gam2}
\end{eqnarray}
Recalling (\ref{SKOangle}) and constant of motion $\varepsilon$ from (\ref{epsil}) we obtain orbit equation for coordinate angle $d\theta$ in form of
\begin{eqnarray}
d\theta &=& \frac{\alpha \; dr}{r^2 \sqrt{\varepsilon^2 - \frac{1}{\gamma^2_g}\left ( 1+ \frac{\alpha^2}{r^2} \right ) }} \label{eq:gam3}
\end{eqnarray}
Above express the same orbit equation that we have from Schwarzschild metric \cite{Schw}.\\
~\\
In \cite{nasz} we have also shown, that we may consider light behavior treating $\gamma_g$ as refracting index for light speed $v_c$ in vicinity of massive object.
\begin{equation}
\gamma_g = \frac{c}{v_c} \label{lsinX}
\end{equation}
From Fermat principle and optics laws \cite{Fermat} we know, that light ray trajectory in the spherical medium with refracting index $\gamma_g(r)$ will be characterized by a constant $\alpha_c$ which corresponds to an angular momentum known from classical mechanics
\begin{equation} 
\alpha_c \equiv  r_{min} \cdot \gamma_g(r_{min})
\end{equation}
where $r_{min}$ is the smallest radius on trajectory.\\
Following \cite{Fermat} we may now note light trajectory in SKO's reference frame as
\begin{equation}
d\phi= \frac{\alpha_c \; dr}{r^2 \sqrt{\gamma_g^2 - \frac{\alpha_c^2}{r^2}}}
\end{equation}
what for coordinate angle $d\theta$ from (\ref{SKOangle}) means
\begin{equation}
d\theta= \frac{\gamma_g\alpha_c \; dr}{r^2 \sqrt{\gamma_g^2 - \frac{\alpha_c^2}{r^2}}} \label{lightTraj}
\end{equation}
Above express exactly the same equation of light trajectory that we obtain from Schwarzschild metric \cite{Schw}.

\subsection{Gravitational effective potential}

Thanks to (\ref{epsil}) and (\ref{angmom}) we will rewrite normalized Lagrangian as:
\begin{equation} 
\ell =  \frac{1}{\varepsilon}\sqrt{1-\frac{r_s}{r}} -  \sqrt{1- \frac{\alpha^2}{r^2 \varepsilon^2\gamma_g^2} - \left ( \frac{dr}{d\tau} \right )^2} \label{lagforR}
\end{equation}
and using Euler-Lagrange equations after simply calculations we obtain radial acceleration in SKO reference frame $a_{daf}$ equal to
\begin{equation}
a_{daf}  \equiv - \nabla \ell = - \frac{r_s}{2r^2 \varepsilon}\gamma_g + \frac{\alpha^2}{r^3 \varepsilon} \gamma_g \left ( 1- \frac{3 r_s}{2r} \right )
\end{equation}
and according to (\ref{lag_red}) after renormalization we obtain the force acting on rest mass, equal to
\begin{equation}
E_0 \varepsilon \cdot a_{daf}  = E_0 \left ( - \frac{r_s}{2r^2 }\gamma_g + \frac{\alpha^2}{r^3 } \gamma_g \left ( 1- \frac{3 r_s}{2r} \right ) \right ) \label{the force}
\end{equation}
As we see, at $r=\frac{3}{2}r_s$ centrifugal force vanishes what gives rise to Black Hole phenomena and defines photon sphere.\\
~\\
We may normalize above force to acceleration $a_{gr}$ acting on rest mass, and calculated in respect to time coordinate t 
\begin{equation}
a_{gr} \equiv \frac{\varepsilon a_{daf}}{\gamma_g} = - \frac{r_s}{2r^2} +  \frac{\alpha^2}{r^3} \left ( 1- \frac{3 r_s}{2r} \right ) \label{acc_coor}
\end{equation}
what let us define effective gravitational potential $V_{gr}$ as
\begin{equation}
V_{gr} \equiv - \int a_{gr} \; dr = - \frac{r_s}{2r} + \frac{\alpha^2}{2 r^2} - \frac{\alpha^2 r_s}{2 r^3} + C \label{effpot}
\end{equation} 
~\\
As we see, obtained acceleration $a_{gr}$ and effective potential $V_{gr}$ are the same, that we obtain from General Relativity for Schwarzschild case \cite{Schw}. 

\subsection{Proper time and energy \label{ptie}}

What still may be confusing, it is test bodies proper time value and relation expressed in (\ref{epsil}). We will consider it, for a moment.\\
~\\
 Let us recall Schwarzschild metric tensor noted as $\varrho$ and invariant of the metric noted as $ds_{\varrho}$
\begin{equation}
\varrho_{\mu\nu} \equiv  \begin{bmatrix}
\frac{1}{\gamma_g^2} & 0 & 0 & 0\\ 
0 & -\gamma_g^2 & 0 & 0\\ 
0 & 0 & -r^2 & 0\\ 
0 & 0 & 0 & -r^2\sin^2(\theta)
\end{bmatrix} 
\end{equation}
\begin{equation}
ds_{\varrho}^2 \equiv \varrho_{\mu\nu} dx^\mu dx^\nu = \frac{dt^2}{\gamma_g^2}-\gamma_g^2dr^2-r^2d\theta^2-r^2\sin^2(\theta)d\varphi^2
\end{equation}
One may easy calculate, that invariant value $ds$ in our metric $g_{\mu\nu}$ (\ref{metTen}) in relation to invariant value $ds_{\varrho}$ in Schwarzschild metric is
\begin{equation}
\frac{ds^2}{d\tau^2}=\frac{ds_{\varrho}^2}{dt^2} + \beta_g^2  \label{noSpin}
\end{equation}
what shows, that in our metric (\ref{metTen}) light speed is equal to $1/\gamma_g$. This weakness of the metric is apparent, and we will explain the real meaning of this property in the next sections. \\
~\\
Let us recall relation (\ref{epsil}), which is valid also in Schwarzschild metric and transform to
\begin{equation}
\frac{1}{\varepsilon^2} = \frac{ds^2}{d\tau^2}\gamma_g^2=\frac{ds_{\varrho}^2}{d\tau^2} + \beta_g^2\gamma_g^2
\end{equation}
We see that $\varepsilon$ indeed represents energy of the body in infinity what agrees with Schwarzschild solution.

\section{Gravitational field properties and origin of SKO's metric tensor \label{Maxwell}}

Now let us explain why above solution works fine, despite that we use different metric than GR and light seems to travel with $\frac{1}{\gamma_g}$ speed.\\
~\\
Let us start from the beginning with already introduced gravitational field scalar $\frac{1}{\gamma_g}$. This scalar field affects only SKO's propertime, and should give metric tensor
\begin{equation}
\hat{g}_{\mu\nu} \equiv  \begin{bmatrix}
\frac{1}{\gamma_g^2} & 0 & 0 & 0\\ 
0 & -1 & 0 & 0\\ 
0 & 0 & -r^2 & 0\\ 
0 & 0 & 0 & -r^2\sin^2(\theta)  \label{metbase}
\end{bmatrix}
\end{equation}
To explain how gravitational field changes above metric into (\ref{metTen}), let us go through gravitational vector fields analyses.

\subsection{Explanation and generalization for other GR solutions}

On flat Minkowski spacetime, if we denote velocity of the body as $\vec{v}$ and its time dilation factor as $\gamma=\frac{dt}{d\tau}$ than from basic calculus rules we may express its material acceleration (in respect to proper time) by local velocity derivatives as 
\begin{equation}
\frac{d\vec{v}}{d\tau} = \gamma \frac{\partial \vec{v}}{\partial t} +  \gamma\vec{v} \cdot \nabla  \vec{v}  \label{propv}
\end{equation}
Now, let us notice, that
\begin{equation}
\nabla \frac{1}{\gamma}  = \nabla \sqrt{1-\vec{v} \cdot \vec{v}} = - \gamma \vec{v} \cdot \nabla \vec{v} 
\end{equation}
therefore we may rewrite (\ref{propv}) as
\begin{equation}
- \frac{d\vec{v}}{d\tau} = \nabla \frac{1}{\gamma}  - \frac{\partial \vec{v}}{\partial \tau}
\end{equation}
We may therefore treat velocity and acceleration as some vector fields. Let us take advantage of this property to consider gravity as field.\\
~\\
At first lest us introduce scalar $\beta_g$  equal to
\begin{equation}
\beta_g (r) \equiv \sqrt{1-\frac{1}{\gamma_g^2}}=\sqrt{\frac{r_s}{r}} \label{intV}
\end{equation} 
As we already know, this scalar field represents move of free-falling surroundings and - by analogy -  illusory move of the SKO against free-falling surroundings.\\
~\\
Now, we define following vector fields 
\begin{eqnarray}
&& \vec{B}\equiv - \beta_g \cdot \vec{\hat{e}_{r}}     \label{eq:Vfield}  \\
&& \vec{G} \equiv \nabla \frac{1}{\gamma_g}  - \frac{\partial \vec{B}}{\partial \tau}  =  -\frac{d\vec{B}}{d\tau} \label{eq:Gfield} \\
&& \vec{\Omega} \equiv \nabla \times \vec{B} \label{eq:rotac} 
\end{eqnarray}
For stationary observer (SKO), field $\vec{B}$ does not change in time, therefore \vec{G} field simplifies to
\begin{equation}
\vec{G} = \nabla \frac{1}{\gamma_g}= \frac{r_s}{2r^2}\gamma_g \cdot \vec{\hat{e}_{r}} 
\end{equation}
giving proper value for gravitational acceleration for stationary SKO.\\
~\\
For moving test particle we would have
\begin{equation}
\frac{\partial \vec{B}}{\partial \tau} = - \left ( \frac{\partial  \beta_g}{\partial \tau} \cdot \vec{\hat{e}_{r}} + \beta_g \frac{\partial  \vec{\hat{e}_{r}}}{\partial \tau} \right ) = \frac{\beta_g}{2r}\frac{\partial r(\tau)}{\partial \tau} \vec{\hat{e}_{r}} - \beta_g \frac{\partial  \vec{\hat{e}_{r}}}{\partial \tau}
\end{equation}
For free-falling particle, gravitational acceleration should vanish, yielding
\begin{equation}
\nabla \frac{1}{\gamma_g} = \frac{\partial \vec{B}}{\partial \tau}
\end{equation}
what requests, that free-falling move with zero angular momentum will be in radial direction with local velocity $\frac{\partial r(\tau)}{\partial \tau} = \beta_g\gamma_g$\\ 
~\\
To derive gravitational acceleration for orbiting particle, let us at first multiply our field with $\frac{1}{\varepsilon}$ obtaining new scalar fields $\gamma_v$ and $\beta_v$ representing spatial velocity and time dilation of the test particle, with proper time of the particle denoted as $ds$
\begin{eqnarray}
&& \frac{1}{\gamma_v} = \frac{1}{\varepsilon} \cdot \frac{1}{\gamma_g} \\
&& \vec{V}\equiv - \sqrt{1-\frac{1}{\gamma_v^2}} \cdot \vec{\hat{e}_v} \\
&& \vec{A} \equiv \nabla \frac{1}{\gamma_v}  - \frac{\partial \vec{V}}{\partial s} = -\frac{d\vec{V}}{ds} \\
\end{eqnarray} 
where $\vec{\hat{e}_v}$ is unit vector that points direction of the particle's velocity.\\
~\\
We already know, from previous sections, that $\varepsilon$ corresponds to conserved energy of the particle, therefore $\vec{V}$ field is indeed velocity of test particle in SKO's reference frame. If we consider test particle as moving with no acceleration (move on geodesics), than it yields 
\begin{equation}
\frac{d\vec{V}}{d s} =0 \;\; \to \;\; \frac{\partial \vec{V}}{\partial s} = \nabla  \frac{1}{\gamma_v} = \frac{1}{\varepsilon} \nabla \frac{1}{\gamma_g} \label{assum}
\end{equation}
Therefore gravitational field for such body could be rewritten, as:
\begin{equation}
\vec{A} = \nabla \left ( \frac{1}{\varepsilon} \frac{1}{\gamma_g} - \frac{1}{\gamma_v} \right) = 0
\end{equation}
Let us introduce our Lagrangian, that corresponds to above
\begin{equation}
\ell = \frac{1}{\varepsilon}\frac{1}{\gamma_g} - \frac{1}{\gamma_v} =  \frac{1}{\varepsilon}\sqrt{1-\vec{\beta_g} \cdot \vec{\beta_g}} - \sqrt{1-\vec{V} \cdot \vec{V}}
\end{equation}
From fundamental lemma of calculus of variations that stands behind Euler-Lagrange equation we know, that for generalized coordinates $x_i$
\begin{equation}
\frac{d}{d\tau}\frac{\partial \frac{1}{\gamma(\dot{x_i},x_i)} }{\partial \dot{x_i}} = \frac{\partial \frac{1}{\gamma(\dot{x_i},x_i)} }{\partial x_i}  \label{ELcor}
\end{equation}
where in our case "dot" represents derivative with respect to SKO's proper time noted as $\tau$.\\
~\\
We know, that $\vec{V}$ and $\vec{\beta_g}\gamma_g$ are functions of $\tau$ then let us rewrite Lagrangian as
\begin{equation}
\ell = \frac{1}{\varepsilon}\sqrt{1-\vec{\beta_g}\gamma_g \cdot \vec{\beta_g}\gamma_g \cdot \frac{1}{\gamma_g^2}} - \sqrt{1-\vec{V} \cdot \vec{V}}
\end{equation} 
Now, we calculate time derivatives
\begin{eqnarray}
&& \frac{d}{d\tau}\frac{\partial \ell}{\partial V} = \frac{d\vec{V}\gamma_v}{d\tau} = \frac{d\vec{V}}{ds} + \vec{V} \frac{d\gamma_g \varepsilon}{d\tau}  \nonumber \\
&& \frac{d}{d\tau}\frac{\partial \ell}{\partial \beta_g\gamma_g} = - \frac{1}{\varepsilon} \frac{d \left ( \vec{\beta_g}\gamma_g^2 \cdot \frac{1}{\gamma_g^2} \right ) }{d\tau} = - \frac{1}{\varepsilon} \frac{d\vec{\beta_g}}{d\tau}
\end{eqnarray}
We know from assumption (\ref{assum}) that $\frac{d\vec{V}}{ds} = 0$\\
We also know, that SKO's time dilation factor does not change in time, since SKO by definition is stationary in respect to gravitational source.\\
Thanks to gravitational acceleration definition (\ref{eq:Gfield}) we therefore obtain
\begin{equation}
\frac{d}{d\tau}\frac{\partial \ell}{\partial \dot{x_i}} = - \frac{1}{\varepsilon} \frac{d\vec{\beta_g}}{d\tau} = \frac{1}{\varepsilon} \frac{d\vec{B}}{d\tau} = - \frac{1}{\varepsilon} \vec{G}
\end{equation}
what thanks to (\ref{ELcor}) let us to define gravitational acceleration as 
\begin{equation}
\vec{G} = -\nabla \ell \cdot \varepsilon  \label{Gfield:eqval}
\end{equation}
As we have already seen in previous sections, gradient of the Lagrangian indeed gives expected gravitational acceleration value. \\
~\\
Since we have derived gravitational field properties and Lagrangian from fundamental properties of the calculus, it will work for all GR solution that conserves energy of the test particle. For other GR solutions (e.g. Kerr solution), the only diffrence wil be in definition of the gravitational dilation factor for SKO. \\
~\\
It means, that gravity may be perceived as field for all GR solutions, that conserves relation (\ref{epsil}). It seems to be the solution for the problem of Gravitomagnetic description of gravity (GEM) considered e.g. in \cite{gem1}, \cite{gem2}, \cite{gem3}, \cite{gem4}.

\subsection{Gravitational waves \label{gwaves}}
We may check, if there are conserved vector field relations, which let us to derive d'Alambertian for the field that would describe gravitational waves.\\
~\\
We may easy check, that
\begin{eqnarray}
\nabla \times \vec{G} = - \frac{\partial \vec{\Omega}}{\partial \tau} = - \frac{\partial (\nabla \times \vec{B})}{\partial \tau} = \nabla \times - \frac{\partial \vec{B}}{\partial \tau} \label{con1}
\end{eqnarray}
what thanks to (\ref{eq:Gfield}) is true.\\
~\\
To consider d'Alambertian for the field we also request
\begin{eqnarray}
\nabla \times \vec{\Omega} = \frac{\partial \vec{G}}{\partial \tau} \label{con2}
\end{eqnarray}
We may transform above into condition for \vec{B} field. Let us define four-potential
\begin{eqnarray}
B^\mu = (\frac{1}{\gamma_g}, \vec{B})
\end{eqnarray}
Next, thanks to vector field transformation properties we may transform (\ref{con2}) into condition
\begin{equation}
\square \vec{B} =  - \nabla \left ( \partial_\mu B^\mu \right ) \label{genConV}
\end{equation}
For Schwarzschild case, $\vec{G}$ does not depend on time, therefore (\ref{genConV}) simplifies to just
\begin{eqnarray}
\nabla (\nabla \cdot \vec{B}) = \nabla^2 \vec{B} \label{eqVar}
\end{eqnarray}
As it is easy to calculate, introduced vector field $\vec{B}$ for Schwarzschild case satisfies this equation giving
\begin{equation}
\nabla (\nabla \cdot \vec{B}) = \nabla^2 \vec{B} = \frac{9}{4r^2}\vec{\beta_g} \label{eqVdiv}
\end{equation}
From above reasoning we see, that if we want to consider gravitational waves, we need time-dependent gravitational acceleration and/or time dependent free-falling velocity. \\
~\\
Such situation for sure will take place for complex gravitational system, consisted of e.g. two co-interacting Schwarzschild sources. Such co-interacting entities would produce around time dependent $\vec{G} and \vec{B}$ fields, that depends of the phase of rotation of these entities.

\subsection{Metric tensor for SKO in DaF framework}

Free-falling test particle in every particular point of its fall, may be considered as stationary SKO observing free-falling gravitational source, thanks to idea of co-moving frames \cite{Rindler} introduced at first in Rindler transformation. \\
~\\
Let us then check, how this field changes in time, if the source of gravity is free-falling onto SKO. We define such free-falling velocity $\vec{v}$, as
\begin{equation}
\vec{v}\equiv \left (\beta_g\gamma_g \vec{\hat{e}_r} \; , \; r \frac{d\theta}{d\tau}\vec{\hat{e}_{\theta}} \; ,\; r \sin{\theta}  \frac{d\varphi}{d\tau} \vec{\hat{e}_{\varphi} } \right )
\end{equation}
Next let us calculate
\begin{equation}
\vec{v} \cdot \nabla \vec{B} = \left ( \vec{G} \; , \;  \frac{d\theta}{d\tau} \beta_g \vec{\hat{e}_{\theta}} \;, \; \sin{\theta}  \frac{d\varphi}{d\tau} \beta_g \vec{\hat{e}_{\varphi} }  \right )
\end{equation}
Velocity gradient is to velocities what the deformation gradient is to displacements. Since $\vec{B}$ represents indeed some velocity, its four-gradient should be composition of some share and rotation, what we observe above. \\
~\\ 
We may separate this two elements by dividing velocity into radial and angular components
\begin{eqnarray}
&& \vec{v_{rad}}\equiv \left (\beta_g\gamma_g \vec{\hat{e}_r} , \; 0, \; 0\right ) \\
&& \vec{v_{ang}} \equiv \left (0 , \; r \frac{d\theta}{d\tau} \vec{\hat{e}_{\theta}} \;, \; r \sin{\theta}  \frac{d\varphi}{d\tau} \vec{\hat{e}_{\varphi} }  \right ) \\
&& \vec{v} = \vec{v_{rad}} + \vec{v_{ang}}
\end{eqnarray}
Let us introduce auxiliary $\vec{\mathbb{S}}$ representing rotation of the frame and equal to
\begin{equation}
\vec{\mathbb{S}} \equiv \vec{v_{ang}} \cdot \nabla \vec{B} = \left ( 0, \;  \frac{d\theta}{d\tau} \beta_g \vec{\hat{e}_{\theta}} \;, \; \sin{\theta}  \frac{d\varphi}{d\tau} \beta_g \vec{\hat{e}_{\varphi} }  \right )  \label{OmegaF}
\end{equation}
We also know from previous sections, that
\begin{equation}
\vec{G} =  \vec{v_{rad}} \cdot \nabla \vec{B} 
\end{equation}
We know from GR that Schwarzschild sources cause the spin of the test bodies exactly with the rate expressed in $\vec{\mathbb{S}}$. Let us therefore treat this quantity as the measure of the spin of the test bodies moving in vicinity of gravitational sources.\\
~\\
To avoid introducing intrinsic spin tensor, since this spin concerns all test bodies, it will be more convenient to consider test particle as rotating along imaginary axis. Such "imaginary axis rotation" of test particle indeed represents the rotation of the gravitational source in test particle reference frame, caused by the test particle spin. \\
~\\
We may perform this operation using present (\ref{metbase}) metric tensor $\hat{g}$ as
\begin{eqnarray}
&& ds^2 = \hat{g}_{\mu\nu}  dx^\mu dx^\nu +  r^2\vec{\mathbb{S}}^2 d\tau^2 = \nonumber \\
&& = dt^2\frac{1}{\gamma_g^2}-dr^2-r^2d\theta^2\frac{1}{\gamma_g^2}-r^2\sin^2(\theta)d\varphi^2\frac{1}{\gamma_g^2} \label{metReal}
\end{eqnarray}
We may then just change metric tensor $\hat{g}_{\mu\nu}$  into $g_{\mu\nu}$ defined in (\ref{metTen}) what will bring the same result. As we have already seen, such approach works fine and reproduces results of the Schwarzschild case.

\subsection{Massless particles in DaF framework}

Spin rate incorporated into the metric in (\ref{metReal}) brings some important consequence for the light and massless particles behavior in DaF framework. \\
~\\
In the (\ref{noSpin}) we have already seen that this additional spin related component is the clue of the difference between propertime in DaF and Schwarzschild metric invariant, resulting with light speed equal to $\frac{1}{\gamma_g}$ in DaF picture.\\
~\\
In our previous article we have also pointed, that if we switch to the gravitational field oriented approach, then we must carefully choose reference frame, because gravity seems to change light speed into $\frac{1}{\gamma_g}$ speed in distant observer reference frame. Now we will solve this enigma.\\
~\\
First easy explanation would be, that for light and massless particles we cannot consider spin, therefore in DaF framework we should make special case for light and consider light with SKO metric tensor $ \hat{g}_{\mu\nu}$ defined in (\ref{metbase}) without spin component. It sounds as the best solution, however light trajectory was correctly calculated for the metric $ g_{\mu\nu}$ with the spin included in the metric tensor as we seen in (\ref{lightTraj}). \\
~\\
We will then choose second explanation. Suppose, that light indeed travels with $c=1$ speed in SKO's reference frame with the metric defined by $ \hat{g}_{\mu\nu}$ giving
\begin{equation}
d\tau^2 = dr^2 + r^2 (d\theta^2 + \sin^2{\theta} d\varphi^2)
\end{equation}
According to (\ref{metReal}) we obtain its proper time coming only from its spin and equal to
\begin{equation}
ds^2 =  r^2(d\theta^2 + \sin^2{\theta} d\phi^2)\beta_g^2 =  (d\tau^2-dr^2)\beta_g^2
\end{equation}
showing, that for massless particles, spin is the same for angular rotation and "radial-to-propertime" axis, exactly as we see in Lorentz transformation expressed as hyperbolic rotation between time and radial axis \cite{Lorentz}.\\ 
~\\
It means, that massless particles indeed travels with c=1 speed in SKO's reference frames, but we also may assign to them propertime and energy that comes from its spin.\\
~\\
This idea opens new area for exploration for origin of elementary particles rest energy as the energy stored in the spin. We have already shown in previous article \cite{nasz}, that rest energy $E_0$ of the massive body seems to be just Planck limit of the energy stored in the spin of massless particle approximated with Taylor series
\begin{equation}
E_0 = \lim_{r \to l_{planck}} E_{planck} (\gamma_g -1) \approx E_{planck} \frac{r_s}{2l_{planck}} = mc^2
\end{equation}
One could also use our research to reconsider electromagnetic field definition. As it is known, considering electromagnetic four-current $A^\mu \equiv (\phi, \vec{A})$ we may make gauge-fixing introducing new scalar field $\tau$ and transforming
\begin{equation}
\phi \; \to \;  \phi + \frac{\partial \tau}{\partial t}     \;\;   \;\;\;\;  \vec{A} \; \to \; \vec{A} - \nabla \tau   \label{gravEM}
\end{equation}
and such transformation does not change electromagnetic field properties. If we consider $\tau$ as proper time, we would obtain similar description of electromagnetism that we present in DaF approach.\\
~\\
Therefore it seems quite easy to create electromagnetic version of DaF framework similar way that we did for gravity, normalizing electromagnetic four-potential to dimensionless four-velocity and based on Planck units.\\
~\\
Our results from section \ref{gwaves} also points, that if we would describe electromagnetic field similar way than we did for gravity, then we should rather treat charged elementary particle (e.g. electron) as complex system, consisted of some elementary entities (e.g. Planck particles) interacting with time independent basic field (e.g. $\sqrt{1-\frac{L_{Planck}}{r}}$). Such complex particle, would produce all around time dependent field depending on the phase of rotation of internal entities, giving rise to the electromagnetic waves around.\\

\section{Conclusions and open issues}

We have shown, that we can successfully reproduce Schwarzschild case results using gravitational field oriented approach, based on DaF framework and generalize it for other GR solutions. It means, that we are able to describe gravity as field and keep full compliance with GR. \\
~\\
What is the most interesting, we have shown, that by incorporating intrinsic spin into the metric, our understanding of the massless particles dynamics may be changed. It should open discussion about energy related to the spin and its consequences for rest energy of the elementary particles.

\end{document}